\documentclass[runningheads]{llncs}
\usepackage{graphicx}

\usepackage{xcolor}
\usepackage{colortbl,booktabs}
\usepackage{graphicx}
\usepackage{epstopdf}
\usepackage{blindtext, graphicx, mathrsfs, amsmath, amsfonts}
\usepackage{booktabs}
\usepackage{multirow, minipage-marginpar}
\usepackage{floatrow}
\usepackage{caption}
\usepackage{subfigure}
\begin{document}
\title{Modality-Pairing Learning for Brain Tumor Segmentation} 
%
%
\author{Yixin Wang\inst{1,2,3} \and
Yao Zhang\inst{1,2,3} \and
Feng Hou\inst{1,2,3} \and
Yang Liu\inst{1,2,3} \and
Jiang Tian\inst{3}\and
Cheng Zhong\inst{3}\and
Yang Zhang\inst{4}\and
Zhiqiang He\inst{1,2,4}}
\authorrunning{Y. Wang et al.}
%
\institute{Institute of Computing Technology, Chinese Academy of Sciences, Beijing, China  \and
University of Chinese Academy of Sciences, Beijing, China
\email{\{wangyixin19,zhangyao215,houfeng19\}@mails.ucas.ac.cn}\\ \and
AI Lab, Lenovo Research, Beijing, China
\email{\{liuyang117,tianjiang1,zhongcheng3\}@lenovo.com}\\ \and
Lenovo Corporate Research \& Development, Lenovo Ltd., Beijing, China \\
\email{\{zhangyang20, hezq\}@lenovo.com}
}
\maketitle              
\begin{abstract}
Automatic brain tumor segmentation from multi-modality Magnetic Resonance Images (MRI) using deep learning methods plays an important role in assisting the diagnosis and treatment of brain tumor. However, previous methods mostly ignore the latent relationship among different modalities. In this work, we propose a novel end-to-end Modality-Pairing learning method for brain tumor segmentation. Paralleled branches are designed to exploit different modality features and a series of layer connections are utilized to capture complex relationships and abundant information among modalities. We also use a consistency loss to minimize the prediction variance between two branches.
Besides, learning rate warmup strategy is adopted to solve the problem of the training instability and early over-fitting. Lastly, we use average ensemble of multiple models and some post-processing techniques to get final results. 
Our method is tested on the BraTS 2020 online testing dataset, obtaining promising segmentation performance, with average dice scores of $0.891, 0.842, 0.816$ for the whole tumor, tumor core and enhancing tumor, respectively. We won the second place of the BraTS 2020 Challenge for the tumor segmentation task.
\end{abstract}
\section{Introduction}
Accurate diagnosis and segmentation of brain tumor are crucial to successful surgery treatment. However, manual annotation requires human experts, which is time-consuming, tedious and expensive. In recent years, motivated by the success of deep learning, researchers have attempted to apply deep learning-based approaches to segment various tumors in medical images. 
Fully convolutional networks (FCN)\cite{fcn}, U-Net\cite{unet} and V-Net\cite{vnet} are popular networks for medical image segmentation. 3D U-Net\cite{3dunet} quickly became the priority choice due to its ability to capture spatial context information. Furthermore, various strategies and optimization processes have also been applied to these networks to achieve higher segmentation precision. 

However, designing a highly-efficient and reliable segmentation algorithm for brain tumor is much more difficult due to the variable size, shape and location of target tissues. What's more, class imbalance is another major challenge since the lesion areas are extremely small and suffer from background domination.
In BraTS 2018, the winner Myronenko et al.\cite{18top1} followed the encoder-decoder structure of CNN and added the variational auto-encoder(VAE) branch to reconstruct the input images jointly with segmentation in order to regularize the shared encoder. In BraTS 2019, Jiang et al.\cite{19top1}, who achieved the best performance on the testing dataset, proposed a two-stage cascaded U-Net to progressively refine the prediction. In addition, Zhao et al.\cite{19top2} introduced a set of heuristics in data processing, model devising and result fusing process, which are combined to boost the overall accuracy of the model.
These methods adopted the input-level fusion, which directly integrated the different modalities of MRI brain images. However, these various modalities are in essence different as they provide with different anatomical and functional information about brain structure and physiopathology. Specifically, BraTS datasets contain four modalities for brain tumor MRI images, which can provide complementary information due to their dependence on various acquisition. Native (T1) and post-contrast T1-weighted (T1ce) yield high contrast between gray and white matter tissues, which highlight the tumor without peritumoral. T2-weighted (T2) and T2 Fluid Attenuated Inversion Recovery (Flair) enhance the image contrast for the whole peritumoral edema.

Inspired by \cite{Scalable,HyperDense-Net,HPN}, we propose a Modality-Pairing network to segment brain tumor substructures. The proposed network consists of paralleled branches, using different modalities as input. The first branch uses Flair and T2 to extract features of the whole tumor, while the second takes T1 and T1ce to learn other tumor representations. These two branches are densely-connected to learn the complementary information effectively. Furthermore, a Modality-Pairing loss is utilized to encourage the consistency between the two sets of high-level feature representations. In addition, a learning rate warmup strategy and an ensemble strategy of multiple models are adopted to improve the segmentation performance. Finally, a post-processing stage is implemented to remove spurious or incoherent segmentation objects. We validate the proposed methods on the BraTS2020 training and validation dataset through qualitative and quantitative analyses. Experimental results show that our method can boost the overall segmentation accuracy.
\section{Methods}
\begin{figure*}[htbp]
\flushleft
\includegraphics[scale=0.42]{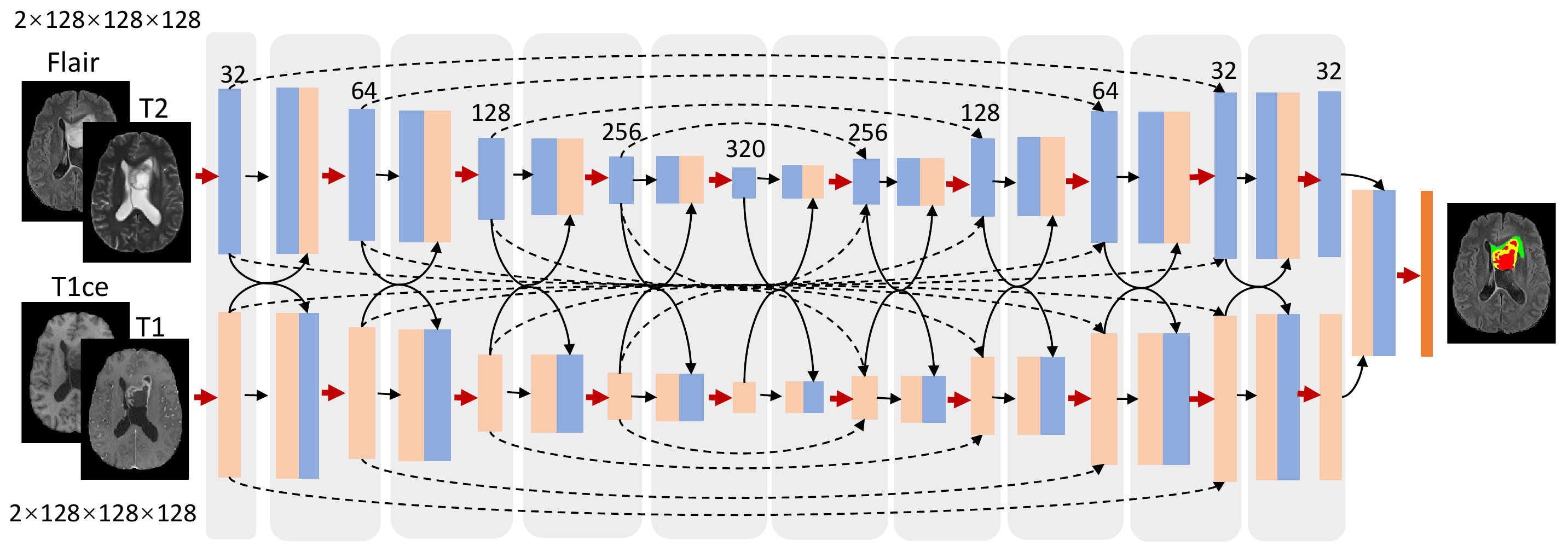}
\caption{ Overview of proposed Modality-Pairing architecture. Each gray region represents a convolutional block. The numbers in the  blocks denote channel numbers. Red arrows correspond to convolutions and dotted arrows indicate skip connections between feature maps. Blue and orange boxes are used to distinguish feature maps from different branches.}
\label{method}
\end{figure*}
\subsubsection{Modality-Pairing Network}
Fig.~\ref{method} shows the architecture of our proposed network. Inspired by the recent works on multi-phase tasks, we present a better way to integrate information from different modalities effectively. Instead of merging the four modalities (T1, T1ce, T2, Flair) at the input of the network, we consider a Modality-Pairing Network which consists of paralleled branches. Each branch focuses on specific modalities and all the branches are properly connected. The purpose of the pairing paths is to derive the features most relevant to each modality and obtain more abundant information among different modalities.
We divide the four modalities into two groups and combine two modalities in each group. Experiments show a better choice to combine T1 and T1ce, T2 and Flair. These two groups are fed into the two branches separately.

All the branches share the same 3D U-Net architecture. It has a U-shape like structure with an encoding and decoding signal path. The encoder includes a series of encoding blocks to contract features, while the decoder is a symmetric path to recover spatial information. The two paths are connected using skip connections to recombine essential high-resolution features. Each block contains two $3\times 3\times 3$ convolutions, each followed by instance normalization and leaky ReLU instead of popular batch normalization and original ReLU. In addition, deep supervision loss is aggregated using multi-level deep supervision output. The output of each deep level is the combination of both branches and computed with corresponding downsampled ground-truth segmentation.

For better fusion of features from different branches, we design a series of connection operation between layers across branches.   
Patches from Flair and T2, T1ce and T1 are concatenated separately to generate two feature maps as input to each branch $A$ and $B$. Let $x_i^m$ be the output of $i^{th}$ layer of branch $m$. This vector can be obtained from the output of the previous layer $x_{i-1}^{m}$ by a mapping $\mathcal{H}(\cdot)$: 
\begin{equation}
\begin{aligned}
x_{i}^{m}=\mathcal{H}\left(x_{i}^{m}\right) &=W * \sigma\left(\mathcal{I}\left(x_{i-1}^{m}\right)\right), i \in\{0, \ldots, l\}.
\end{aligned} 
\end{equation}
The encoder layers can be defined as:
\begin{equation}
x_{i+1}^A =\mathcal{H}\left(\left[x_{i}^A, x_{i}^B \right]\right),\\
x_{i+1}^B =\mathcal{H}\left(\left[x_{i}^B, x_{i}^A \right]\right),
\end{equation}
where $W$ represents the weight matrix, $*$ denotes convolution operation, $\mathcal{I}(\cdot)$ and $\sigma (\cdot)$ represent instance normalization and leaky ReLU, respectively. 

Similar to the encoder, the decoder comprises connections across branches. Besides, the original skip-connections are extended to multi-modality as follows:
\begin{equation}
\begin{aligned}
x_{i}^{\mathrm{A}} &=\mathcal{H}\left(\left[x_{i-1}^{\mathrm{A}}, x_{l-i+1}^{\mathrm{A}}, x_{l-i+1}^B  \right]\right),\\
x_{i}^{\mathrm{B}} &=\mathcal{H}\left(\left[x_{i-1}^{\mathrm{B}}, x_{l-i+1}^{\mathrm{B}}, x_{l-i+1}^{\mathrm{A}}  \right]\right).
\end{aligned}
\end{equation}
Before the final classification layer, feature maps ${\mathrm{X}}^{\mathrm{A}},{\mathrm{X}}^{\mathrm{B}}$ from two branches are fused to generate a final feature map and then fed into a 4-class softmax classifier(i.e. background, ED, NET, ET). In the multi-modality connection settings, our model can yield more powerful feature representations within and in-between different modalities through different branches.  
\subsubsection{Loss}
Dice Loss and Cross Entropy Loss are effective for medical segmentation tasks. In our model, however, different modalities are separately utilized to extract feature maps, which may lead to prediction variance between two branches. In order to better handle this problem, following \cite{HPN}, we adopt a Modality Pairing loss to minimize the distance between the two sets of high-level semantic features. Given two feature maps ${\mathrm{A},\mathrm{B}}$ from two branches, we aim to exploit the consistency between them:
\begin{equation}\small
\mathcal{L}_{\text {MP}}\left(\mathrm{X}_{i}^{\mathrm{A}}, \mathrm{X}_{i}^{\mathrm{B}}\right)=-\frac{\sum_{j=1}^{N}\left(\mathrm{X}_{i j}^{\mathrm{A}}-\overline{\mathrm{X}_{i}^{\mathrm{A}}}\right)\left(\mathrm{X}_{i j}^{\mathrm{B}}-\overline{\mathrm{X}_{i}^{\mathrm{B}}}\right)}{\sqrt{\sum_{j=1}^{N}\left(\mathrm{X}_{i j}^{\mathrm{A}}-\overline{\mathrm{X}_{i}^{\mathrm{A}}}\right)^{2} \sum_{j=1}^{N}\left(\mathrm{X}_{i j}^{\mathrm{B}}-\overline{\mathrm{X}_{i}^{\mathrm{B}}}\right)^{2}}},
\end{equation}
where $N$ is the total number of voxels and ${X}_{i j}$ represent $j^{th}$ voxel in $i^{th}$ sample. 
Based on the above statement, the total loss can be given as follows: 
\begin{equation}
L=\lambda_{1} \cdot L_{Dice}+\lambda_{2} \cdot L_{CE}+\lambda_{3} \cdot L_{MP},
\end{equation}
where $L_{Dice}$, $L_{CE}$, $L_{MP}$ denote the average Dice Loss, Cross Entropy Loss and Modality-Pairing Loss, respectively.  $\lambda_{1}, \lambda_{2}, \lambda_{3}$ are three loss weights which are set as $1,1,0.5$ based on our Dice results of 5-fold cross validation on the training dataset. Meanwhile, deep supervisions are introduced to add auxiliary outputs from decoder layers for better gradient propagation. This auxiliary deep supervision loss only uses $L_{Dice}$ and $L_{CE}$, with $\lambda_{1} = \lambda_{2} = 1$. As a result, the whole network training is optimized by minimizing the loss from both main branches and the auxiliary loss functions.
\subsubsection{Learning Rate Strategy}
Learning rate warmup\cite{warmup} is adopted in our training strategy, where we start training with a much smaller learning rate and then increase it over a few epochs until it reaches our set `initial' learning rate. The effect of learning rate warmup is to prevent deeper layers from training instability and `early over-fitting'. In detail, we start from a learning rate $\eta_{min}$, with $\eta_{min}=0.0005$, and then increase it by a constant amount $0.0005$ at every epoch until it reaches $\eta=\eta_{max}$ with $\eta_{max}=0.01$. This warmup strategy is followed up by a poly learning rate policy$\left(1-\text { epoch } / \text { epoch }_{\max }\right)^{0.9}$.
\subsubsection{Post-processing}
In order to remove spurious or incoherent object, we make connect component analysis and delete any small component (voxels$<$10). What's more, it is noted that there exists no enhancing tumor region in some LGG cases. Therefore, we empirically replace the enhancing tumor regions with less than a threshold ($500$ voxels) by necrosis. 
\subsubsection{Ensemble of Multiple Models}
Due to the high variance of single deep learning models, we adopt model ensemble strategy to combine the segmentation predictions from trained single models. For	simple models, the average of models has greater capacity than single models, which can reduce bias substantially. Specifically, at the end of the inference on the validation dataset, the predicted probability distributions from each single model are averaged as $y_{en}^{M}(x)$:
\begin{equation}
y_{en}^{M}(x)=\frac{1}{M} \sum_{m=1}^{M} y_{m}(x),
\end{equation}
where $y_{m}(x)$ denotes the output probability of model $m \in \{1,...,M\}$ at voxel $x$. Then, each voxel is assigned by the label with the highest probability. 
\section{Experiments}
\subsection{Dataset}
BraTS2020 training dataset\cite{dataset1,dataset2,dataset3,dataset4,dataset5} consists of $369$ multi-contrast MRI scans, out of which $293$ have been acquired from glioblastoma (GBM/HGG) and $76$ from lower grade glioma (LGG). All the multi-modalilty scans contain four modalities: a) native (T1), b) post-contrast T1-weighted (T1Gd), c) T2-weighted (T2), and d) T2 Fluid Attenuated Inversion Recovery (T2-FLAIR) volumes, which are acquired with different clinical protocols and various scanners from multiple (n=19) institutions. Each of these modalities captures different properties of brain tumor subregions: GD-enhancing tumor (ET — label 4), the peritumoral edema (ED — label 2), and the necrotic and non-enhancing tumor core (NCR/NET — label 1).
\subsection{Experimental Settings}
All the experiments are implemented in Pytorch and trained on NVIDIA Tesla V100 32GB GPU. 5-fold cross validation is adopted while training models on the training dataset.
We do data pre-processing following nnU-Net\cite{nnUnet}. Due to the large size of input image, the input patch size is set as $128\times 128\times 128$ and batch size as $2$. Stochastic gradient descent optimizer (SGD) with an initial learning rate of 0.01, a nesterov momentum
of 0.99 and weight decay are used. The maximum number of training iterations is set to 1000 epochs with 20 epochs of linear warmup. The whole model is trained in an end-to-end manner.
\subsection{Evaluation Metrics}
Consistent with the BraTS challenge, we adopt four evaluation metrics. `Dice' measures volumetric overlap between segmentation results and annotations, while `Hausdorff distance (HD95)' measures the $95^{th}$ percentile of values in the set of closet distances between two surfaces. The diagnostic test accuracy `Sensitivity' and `Specificity' are also considered to determine potential over- or under-segmentations. `Sensitivity' shows the percentage of positive instances correctly identified positive and `Specificity' calculates the proportion of actual negatives that are correctly identified.
\subsection{Experimental Results}
The prediction of four subregions is aggregated to generate the whole tumor(WT), tumor core(TC) and enhancing tumor(ET). We first train and test our proposed Modality-Pairing model on BraTS2020 training datasets. Then, the best models and an ensemble model are chosen on BraTS2020 validation dataset.
\subsubsection{BraTS2020 Training Dataset}
\begin{figure*}[htbp]
\flushleft
\includegraphics[scale=0.65]{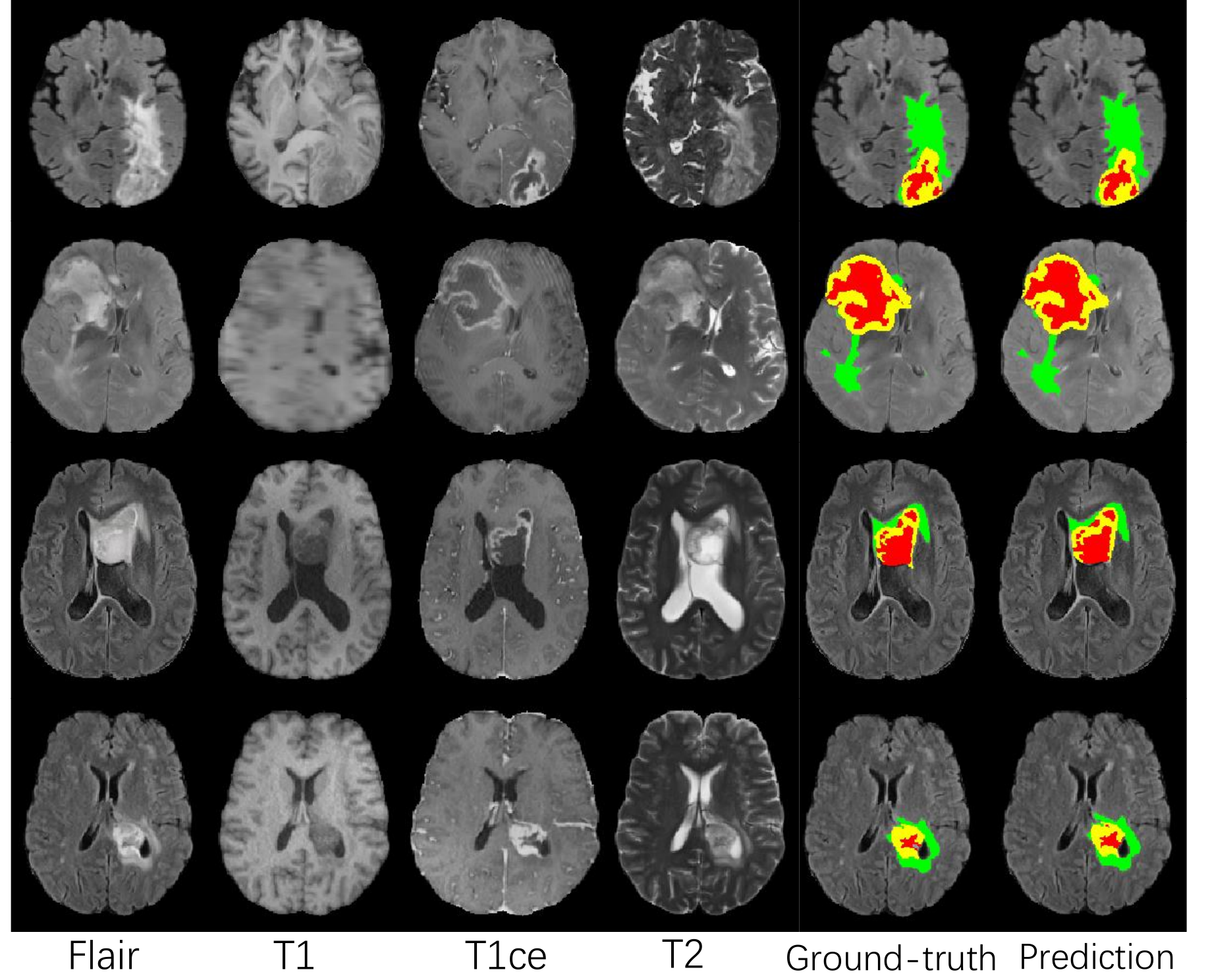}
\caption{Visual segmentation results of proposed Modality-Pairing method. From left to right, show the axial slice of MRI images in four modalities, ground-truth and predicted results. The labels include enhancing tumor (yellow), edema (green)
and necrotic and non-enhancing tumor (red).}
\label{exp}
\end{figure*}

For BraTS2020 training dataset, we use 5-fold cross validation based on a random split manner. We report the result of best single model in Table~\ref{training_dataset}. Compared with Vanilla U-Net, proposed Modality-Pairing method improves the segmentation results of enhancing tumor greatly.
Additionally, Fig.~\ref{exp} also shows some examples of segmentation results of the proposed Modality-Pairing method. As can be seen, the segmentation results are sensibly similar to ground-truth with accurate boundaries and some minor tumor areas identified.
\begin{table}
\caption{The segmentation results of best single model on the training dataset.}\label{training_dataset}
\begin{tabular}{c|c|c|c|c|c}
\toprule
Method & Tumor & Dice & Sensitivity & Specificity & Hausdorff95 \\
\hline
\multirow{3}{*}{\begin{tabular}[c]{@{}c@{}} \\Vanilla\\U-Net\end{tabular}}
&Enhancing Tumor &\ 0.848\  & 0.863 & 1.000 & 12.145 \\
&Whole Tumor &\ 0.923\  & 0.909 & 0.999 & 4.508 \\
&Tumor Core &\ 0.900\  & 0.880 & 1.000 & 3.368 \\
\hline
\multirow{3}{*}{\begin{tabular}[c]{@{}c@{}} \\Modality-Pairing\\learning\end{tabular}}
&Enhancing Tumor &\ 0.863\  & 0.875 & 1.000 & 7.179 \\
&Whole Tumor &\ 0.924\  & 0.910 & 0.999 & 4.131 \\
&Tumor Core &\ 0.898\  & 0.877 & 1.000 & 3.448 \\

\bottomrule
\end{tabular}
\end{table}
\begin{table}
\renewcommand{\arraystretch}{1.3}
\renewcommand{\multirowsetup}{\centering}
\caption{The segmentation results of single Modality-Pairing model on the validation dataset.}\label{val_dataset_singlemodel}
\centering{
\begin{tabular}{c|ccc|ccc|ccc|ccc}
\toprule
     &\multicolumn{3}{c|}{Dice} 
     & \multicolumn{3}{c|}{Sensitivity}
     & \multicolumn {3}{c|}{Specificity}
     & \multicolumn {3}{c}{Hausdorff95}\\ 
\cline{2-13}
     & ET & WT & TC& ET & WT & TC & ET & WT & TC &ET & WT & TC \\
\hline
Mean	& 0.785 &0.907 	&0.837 	&0.783 	&0.901 	&0.804 	&1.000	&0.999	&1.000	&32.25 	&4.39 	&8.34\\ 
StdDev	&0.272 	&0.072 	&0.178 	&0.286 	&0.103 	&0.211 	&0.000	&0.001	&0.000	&101.03 	&5.97 	&33.64 \\
Median	&0.880 	&0.931 	&0.902 	&0.886 	&0.931 	&0.891 	&1.000	&1.000	&1.000	&1.73 	&2.83 	&3.16 \\
25quantile	&0.790 	&0.890 	&0.808 	&0.795 	&0.885 	&0.733 	&1.000	&1.000	&1.000	&1.00 	&1.73 	&1.41 \\
75quantile	&0.923 	&0.949 	&0.944 	&0.951 	&0.964 	&0.943 	&1.000	&1.000	&1.000	&3.00 	&4.24 	&5.83 \\

\bottomrule
\end{tabular}}
\end{table}
\subsubsection{BraTS2020 Validation Dataset}
\begin{figure}[htbp]
	\centering
	\subfigure{
		\begin{minipage}[t]{0.5\linewidth}
			\centering
			\includegraphics[width=2.5in]{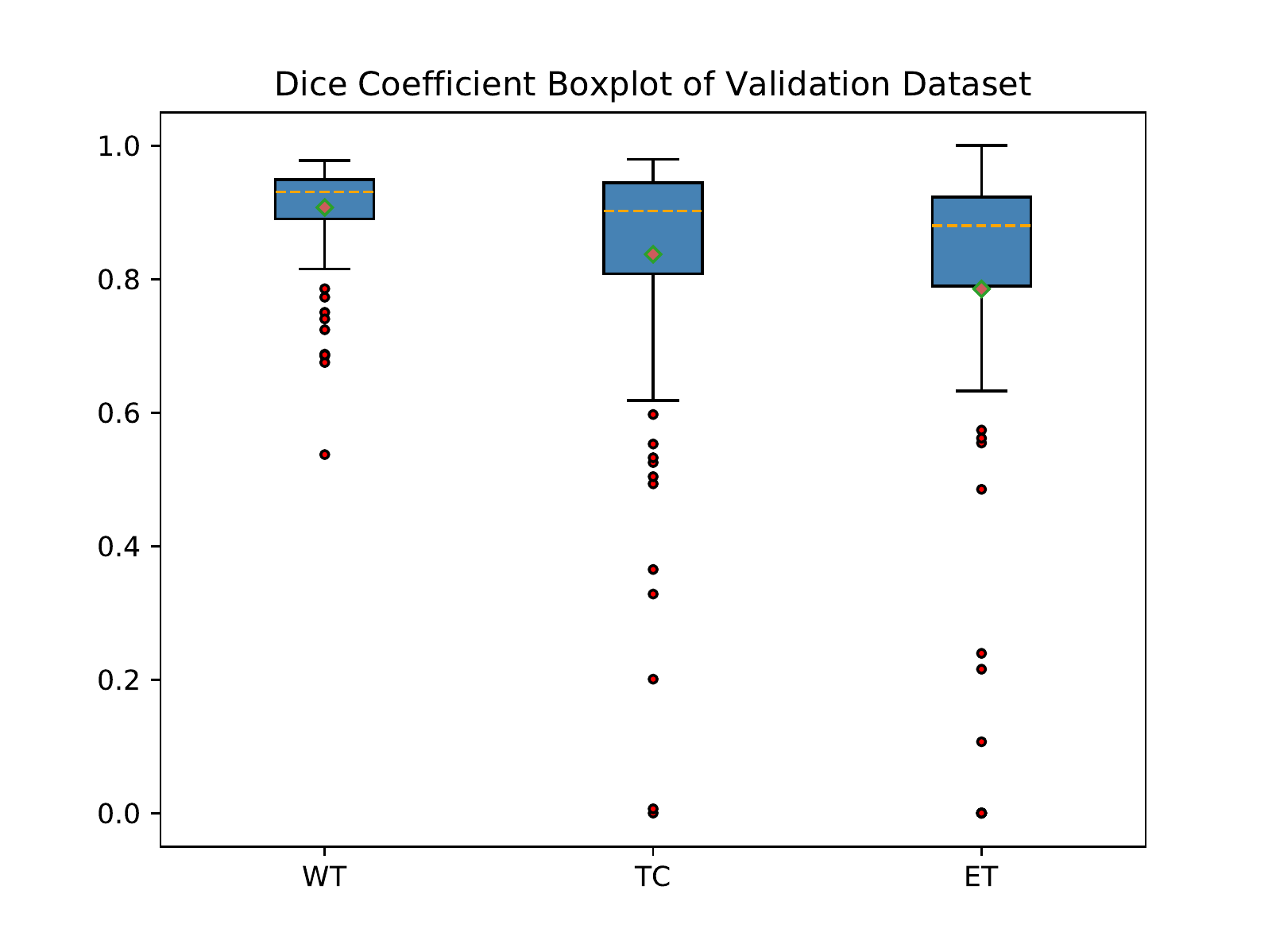}
			\label{dice_boxplot}
		\end{minipage}%
	}%
	\subfigure{
		\begin{minipage}[t]{0.5\linewidth}
			\centering
			\includegraphics[width=2.5in]{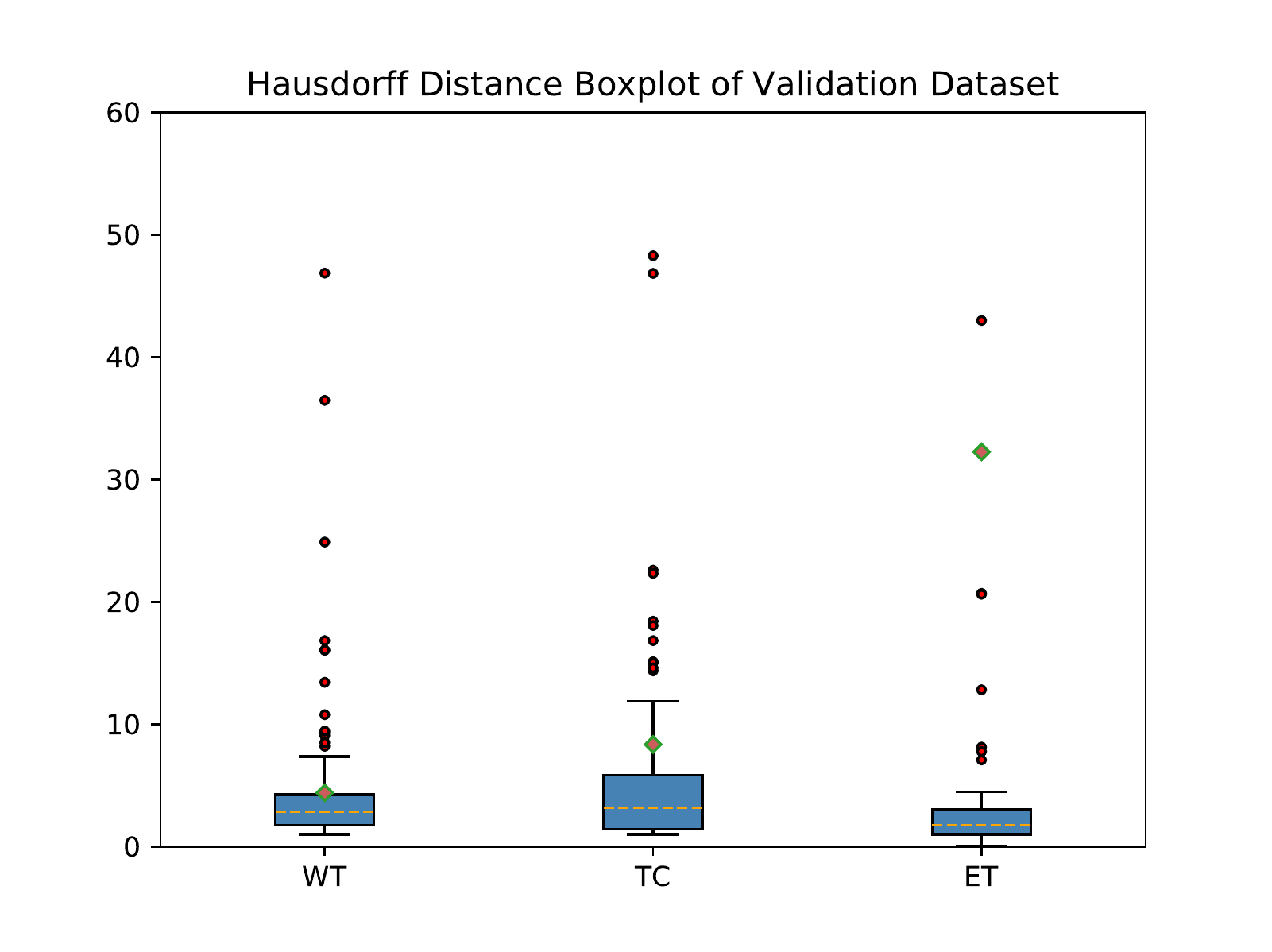}
			\label{HD95_boxplot}
		\end{minipage}%
	}%
	
	\subfigure{
			\begin{minipage}[t]{0.5\linewidth}
				\centering
				\includegraphics[width=2.5in]{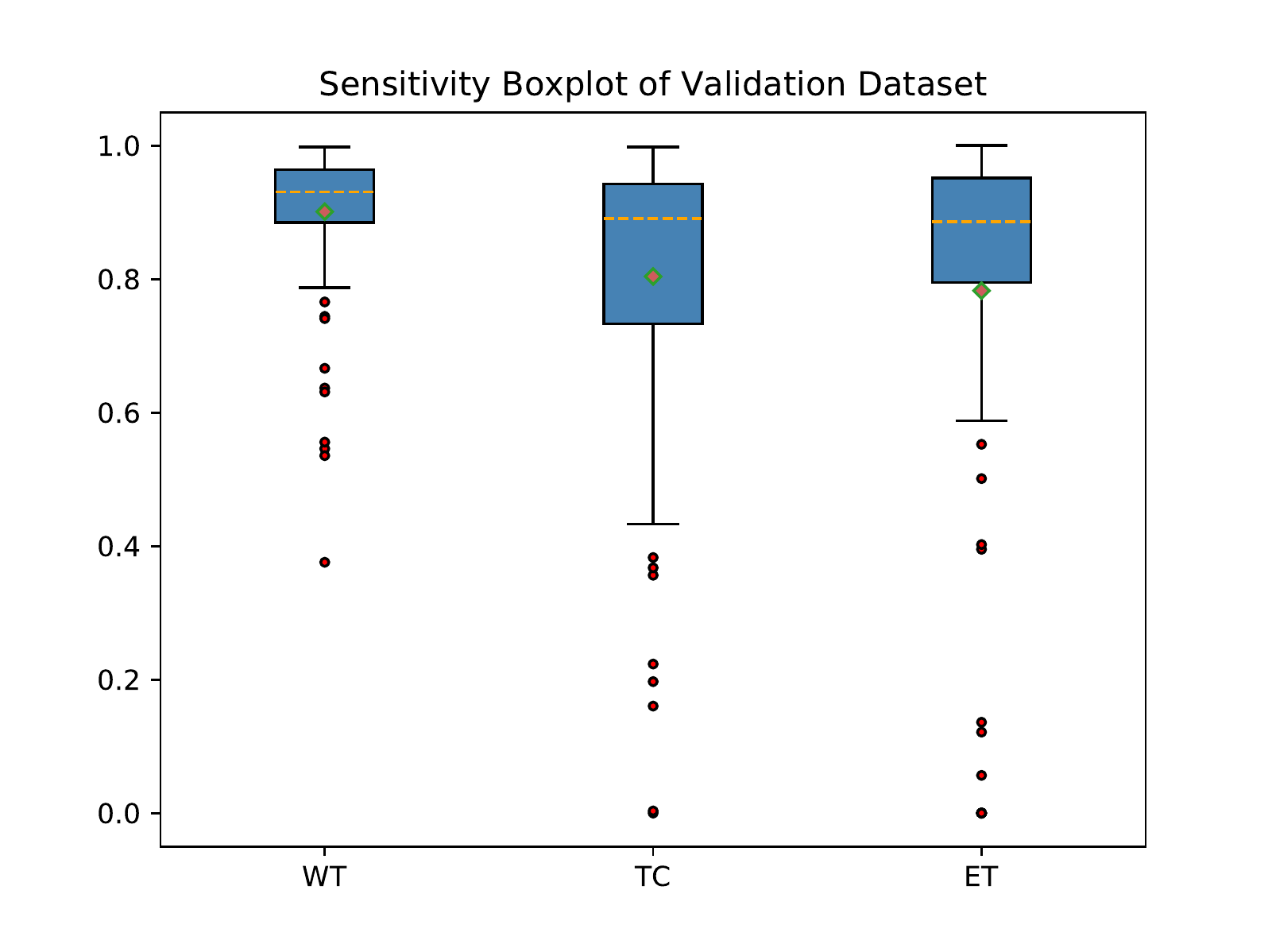}
				\label{Sensitivity_boxplot}
			\end{minipage}%
		}%
		\subfigure{
			\begin{minipage}[t]{0.5\linewidth}
				\centering
				\includegraphics[width=2.5in]{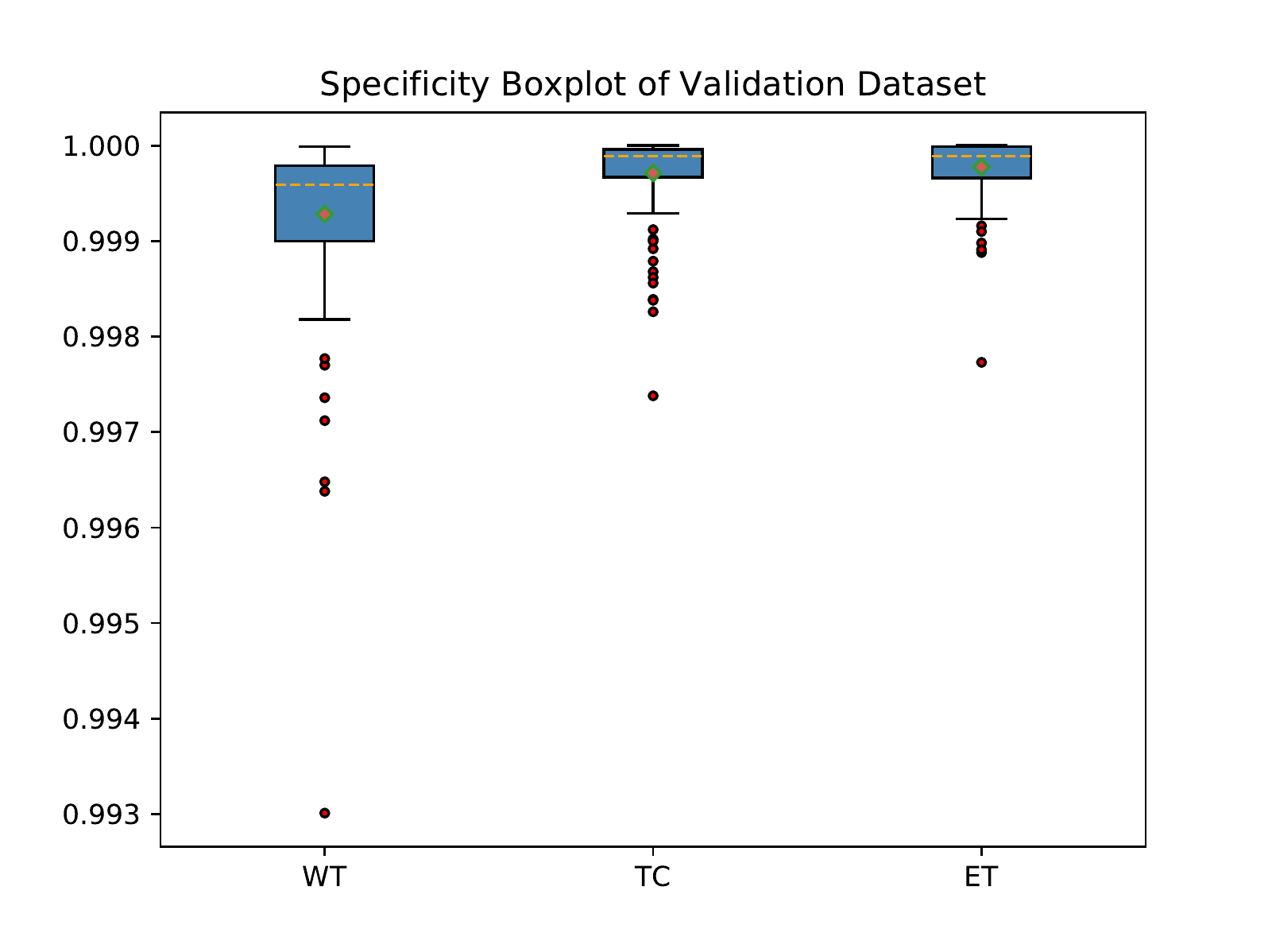}
				\label{Specificity_boxplot}
			\end{minipage}%
		}%
	\centering
	\caption{Box plots of the Dice, Hausdorff95, Sensitivity and Specificity metrics for single best model on the validation dataset evaluated on the Whole Tumor (WT), Enhancing Tumor(ET), and Tumor Core (TC) regions.}
	\label{boxplot}
\end{figure}

We report the results of our proposed model on the validation dataset, consisting of 125 cases with no ground-truth segmentation mask. All the results are evaluated by the official competition platform (CBICA IPP\footnote{https://ipp.cbica.upenn.edu.}).
We first evaluate the performance of each single model and present the best results in Table~\ref{val_dataset_singlemodel}. Fig.~\ref{boxplot} also shows the box plots of segmentation accuracy. According to the Dice and Hausdorff95 metrics, among the three subregions, WT achieves the best performance compared to TC and ET. ET is much more difficult to be detected in some cases with a much higher standard deviation. 
In the ensemble, we choose the top three Modality-Pairing single models from the 5 folds based on the performance on the validation dataset and three Vanilla U-Net single models with proposed learning rate strategy. Table~\ref{val_dataset_ensemble} shows the segmentation results after applying our ensemble method. It can be clearly seen that the ensemble strategy improves the Dice results of enhancing tumor (ET) from 0.785 to 0.793, tumor core (TC) from 0.837 to 0.850, performing better than single best models, which shows the effectiveness of the average ensemble strategy. 
\begin{table}
\renewcommand{\arraystretch}{1.3}
\renewcommand{\multirowsetup}{\centering}
\caption{The segmentation results of ensemble model on the validation dataset}\label{val_dataset_ensemble}
\centering{
\begin{tabular}{c|ccc|ccc|ccc|ccc}
\toprule
     &\multicolumn{3}{c|}{Dice} 
     & \multicolumn{3}{c|}{Sensitivity}
     & \multicolumn {3}{c|}{Specificity}
     & \multicolumn {3}{c}{Hausdorff95}\\ 
\cline{2-13}
     & ET & WT & TC& ET & WT & TC & ET & WT & TC &ET & WT & TC \\
\hline
Mean	&0.787 	&0.908 	&0.856 	&0.786 	&0.905 	&0.822 	&1.000 	&0.999 	&1.000 	&35.01 	&4.71 	&5.70 \\
StdDev	&0.276 	&0.078 	&0.130 	&0.289 	&0.103 	&0.175 	&0.000 	&0.001 	&0.000 	&105.54 	&7.62 	&10.17 \\
Median	&0.885 	&0.931 	&0.907 	&0.887 	&0.934 	&0.895 	&1.000 	&1.000 	&1.000 	&1.73 	&2.83 	&3.00 \\
25quantile	&0.809 	&0.892 	&0.827 	&0.801 	&0.887 	&0.768 	&1.000 	&0.999 	&1.000 	&1.00 	&1.73 	&1.41 \\
75quantile	&0.925 	&0.950 	&0.944 	&0.960 	&0.967 	&0.947 	&1.000 	&1.000 	&1.000 	&2.83 	&4.47 	&5.39 \\

\bottomrule
\end{tabular}}
\end{table}

\subsubsection{BraTS2020 Testing Dataset}
The ensemble strategy was used as the final method to BraTS2020 challenge, with the same ensemble model in the validation phase. Table \ref{testing_data_result} shows our final results on the testing dataset(166 cases), which is provided by the challenge organizer. Our team achieved the second place out of all 78 participating teams. 
\begin{table}
\renewcommand{\arraystretch}{1.3}
\renewcommand{\multirowsetup}{\centering}
\caption{The segmentation results of ensemble model on the testing dataset}\label{testing_data_result}
\centering{
\begin{tabular}{c|ccc|ccc|ccc|ccc}
\toprule
     &\multicolumn{3}{c|}{Dice} 
     & \multicolumn{3}{c|}{Sensitivity}
     & \multicolumn {3}{c|}{Specificity}
     & \multicolumn {3}{c}{Hausdorff95}\\ 
\cline{2-13}
     & ET & WT & TC& ET & WT & TC & ET & WT & TC &ET & WT & TC \\
\hline
Mean	&0.816 	&0.891 	&0.842 	&0.847 	&0.911 	&0.853 	&1.000 	&0.999 	&1.000 	&17.79 	&6.24 	&19.54 \\ 
StdDev	&0.197 	&0.112 	&0.244 	&0.211 	&0.118 	&0.225 	&0.000 	&0.001 	&0.001 	&74.87 	&28.98 	&74.78\\ 
Median	&0.857 	&0.925 	&0.925 	&0.916 	&0.941 	&0.931 	&1.000 	&0.999 	&1.000 	&1.41 	&2.83 	&2.00 \\ 
25quantile	&0.788 	&0.884 	&0.865 	&0.823 	&0.905 	&0.854 	&1.000 	&0.999 	&1.000 	&1.00 	&1.41 	&1.41 \\ 
75quantile	&0.921 	&0.950 	&0.959 	&0.963 	&0.968 	&0.966 	&1.000 	&1.000 	&1.000 	&2.24 	&4.66 	&3.74 \\

\bottomrule
\end{tabular}}
\end{table}
\section{Discussion and Conclusion}
In this paper, we propose a Modality-Pairing learning method using 3D U-Net as backbone network, which exploits a better way to fuse the four modalities of MRI brain images to get compromise for a precise segmentation. This method utilizes paralleled branches to separately extract feature from different modalities and combines them via effective layer connections. On the BraTS2020 online testing dataset, our method achieves average Dice scores of $0.891, 0.842, 0.816$ for the whole tumor, tumor core and enhancing tumor, respectively. The approach won the second
place in the BraTS 2020 challenge segmentation task, with 78 teams participating in the challenge.
%
%
%
\bibliographystyle{splncs04}
\bibliography{mybib}

\end{document}